\documentclass{article}
\usepackage{emulateapj}
\usepackage{psfig}

\newcommand{\FeII}{\mbox{$[\rm FeII]$}}
\newcommand{\PA}{\mbox{$\rm Pa\alpha$}}

\lefthead{Marconi et al.}
\righthead{Active nucleus of Centaurus A}

\begin{document}

\title{Unveiling the Active Nucleus of Centaurus A\altaffilmark{1}}

\altaffiltext{1}{Based  on observations
with the NASA/ESA {\it Hubble Space Telescope}, obtained at the Space
Telescope Science Institute, which is operated by AURA, Inc., under
NASA contract NAS 5-26555}

\author{Alessandro Marconi}
\affil{Osservatorio Astrofisico di Arcetri\\
       Largo E. Fermi 5,\\
       I-50125 Firenze, ITALY\\
       e-mail: marconi@arcetri.astro.it}
       
\author{Ethan J. Schreier, Anton Koekemoer}
\affil{Space Telescope Science Institute\\
       3700 San Martin Drive,\\
       Baltimore, MD 21218, USA\\
       e-mail: schreier@stsci.edu }
       
\author{Alessandro Capetti}
\affil{Osservatorio Astronomico di Torino\\
       Strada Osservatorio 20,\\
       I-10025 Pino Torinese, ITALY\\
       e-mail: capetti@to.astro.it }       
	
\author{David Axon}
\affil{Department of Physical Sciences\\
       University of Hertfordshire\\
       College Lane,\\ 
       Hatfield, Herts AL10 9AB, UK\\
       e-mail:dja@star.herts.ac.uk}
       
\author{Duccio Macchetto\altaffilmark{2}}
\affil{Space Telescope Science Institute\\
       3700 San Martin Drive,\\
       Baltimore, MD 21218, USA\\
       e-mail: macchetto@stsci.edu}
\and       
\author{Nicola Caon}
\affil{Instituto de Astrofisica de Canarias\\
       c/ via Lactea s/n,\\
       E-38200 La Laguna, Tenerife, SPAIN\\
       e-mail: ncaon@ll.iac.es }

\altaffiltext{2}{Affiliated to ESA scientific division}

\authoremail{marconi@arcetri.astro.it}

\begin{abstract}

We report new HST {\it WFPC2} and {\it NICMOS} observations of the center
of the nearest radio galaxy Centaurus A (NGC 5128) and discuss their
implications for our understanding of the active nucleus and jet.
We detect the active nucleus in the near--IR (K and H) and, for the first
time, in the optical (I and V), deriving the spectral energy distribution
of the nucleus from the radio to X-rays. 
The optical and part of the near-IR emission can be explained by
the extrapolation of the X-ray power law reddened by $A_V\sim 14$,
a value consistent with other independent estimates.

The 20pc-scale nuclear disk discovered by Schreier et al.
(\cite{schreier98}) is detected in the $\FeII\lambda 1.64\mu$m
line and presents a morphology similar to that observed in \PA\
with a \FeII/\PA\ ratio typical of low ionization Seyfert 
galaxies and LINERs.
NICMOS 3 Pa$\alpha$ observations in a
50\arcsec$\times$50\arcsec\ circumnuclear region suggest enhanced star
formation ($\sim 0.3$M$_\odot$ yr$^{-1}$)
at the edges of the putative bar seen with ISO, perhaps due to shocks
driven into the gas. 
 
The light profile, reconstructed from V, H and K observations,
shows that Centaurus A has a core profile with a
resolved break at $\sim 4\arcsec$ and suggests a 
black--hole mass of $\sim 10^9$M$_\odot$.
A linear blue structure aligned with the radio/X--ray jet may
indicate a channel of relatively low reddening in which dust has been swept 
away by the jet.

\end{abstract}


\vskip 0.5cm
\section{Introduction}

Centaurus A (NGC5128), the closest giant elliptical galaxy hosting an
active galactic nucleus (AGN) and a jet, is an ideal benchmark for 
testing and addressing some open issues in the unified model of AGNs.
Its relative proximity ($D\sim3.5$ Mpc, Hui et al. \cite{hui})
offers a unique opportunity to investigate the putative supermassive
black-hole, the associated accretion disk and jet, i.e. the entities
which the unified model considers at the core of AGN activity (see 
Wills \cite{wills} for a recent review).
The detailed structure of the galaxy and the stellar and globular cluster
populations can be linked to study the relationship between
star formation and nuclear activity; it should be possible to assess
the role of the merger in the evolution of the giant 
elliptical galaxy, its multiple populations and the triggering and fueling of 
the AGN itself.  However, the study of
this nearest giant elliptical at intermediate wavelengths has been
severely hindered by the presence of a dust lane
which dominates ground-based optical and near-IR observations of the nuclear region.
The dust lane, which obscures the inner half kiloparsec of the
galaxy, with associated gas, young stars
and HII regions, is interpreted as the result of a 
relatively recent merger event between a giant elliptical galaxy and 
a small, gas rich, disk galaxy (Baade \& Minkowski \cite{baade},
Graham \cite{graham},
Malin, Quinn \& Graham \cite{malin}).
See Israel (\cite{israel}) for a recent comprehensive review
on Centaurus A.
\begin{figure*}
\caption{\label{fig:f814w}
Grayscale representation of the mosaic in the WFPC2 F814W
filter. Surface brightness ranges from 0 (white) to 1.6 in units
of $10^{-16}$ erg s$^{-1}$ cm$^{-2}$ \AA$^{-1}$
arcsec$^{-2}$.
Image sizes are $225\arcsec\times 170\arcsec$. North is up and East is left.}
\end{figure*}
\begin{figure*}
\caption{\label{fig:f555w}
Grayscale representation of the mosaic in the WFPC2 F555W
filter. Surface brightness ranges from 0 (white) to 1.2 in units
of $10^{-16}$ erg s$^{-1}$ cm$^{-2}$ \AA$^{-1}$
arcsec$^{-2}$. Sizes and orientation
of image are as in Fig. 1.}
\end{figure*}

The presence of the AGN was revealed and studied through
its radio and X-ray manifestations.
Centaurus  A is a  giant double-lobed radio source, first discovered
by Bolton, Stanley \& Slee (\cite{bolton}) and
it is considered a prototype low luminosity Fanaroff-Riley Class I 
radio galaxy. Its nucleus was found to be a source of
X-ray emission (Bowyer et al. \cite{bowyer}, Kellogg et al. \cite{kellogg},
Grindlay et al. \cite{grindlay75}).  A strong jet was
discovered in the X-rays  (Schreier et al. \cite{schreier79};
Feigelson et al. \cite{feigelson}) and 
subsequently studied in the radio with the VLA (Schreier, Burns \&
Feigelson \cite{schreier81}; Burns, Feigelson \& Schreier \cite{burns};
Clarke, Burns \& Feigelson \cite{clarke86};
Clarke, Burns \& Norman \cite{clarke92})
and VLBI (cf. Jones et al. \cite{jones}); it is the
nearest extragalactic jet.  

In the outer regions, ignoring the dust lane, Centaurus A is a fairly normal
giant elliptical galaxy.  Most of the light comes from an older stellar
population, and the light distribution appears to follow a de Vaucouleurs
law (cf. Graham \cite{graham}).  The extensive
system of shells within the extended elliptical component of the galaxy
(cf. Malin, Quinn \& Graham \cite{malin}) provide further evidence for a merger
event occurring $\lesssim 5-10\times 10^8$ yr ago.

\begin{figure*}
\caption{\label{fig:f336w}
Grayscale representation of the mosaic in the WFPC2 F336W
filter. Surface brightness ranges from 0 (white) to 0.3 in units
of $10^{-16}$ erg s$^{-1}$ cm$^{-2}$ \AA$^{-1}$ arcsec$^{-2}$.
Sizes and orientation of image are as in Fig. 1.}
\end{figure*}
\begin{figure*}
\caption{\label{fig:truecolor}
True color RGB (Red=F814W, Green=F555W and Blue=F336W)
representation of the WFPC2 mosaic. Cutoff values for the images are
as in Fig. 1, 2 and 3. Sizes and orientation of image are as in Fig. 1.}
\end{figure*}
\begin{deluxetable}{lcr}
\tablewidth{0pt}
\tablecaption{\label{tab:logobs}  Observations Log.}
\tablehead{
\colhead{Dataset} &
\colhead{Filter} & \colhead{T$_{exp}$ (sec)}}
\footnotesize
\startdata
\multicolumn{3}{c}{WFPC2 Aug. 1 1997} \nl
U4100101M & F336W  & 1800 \nl
U4100102M & F336W  & 600 \nl
U4100103M & F336W  & 1800 \nl
U4100104M & F336W  & 600 \nl
U4100105M & F555W  & 300 \nl
U4100106M & F555W  & 1000 \nl
U4100107M & F555W  & 1000 \nl
U4100108M & F814W  & 300 \nl
U4100109M & F814W  & 1000 \nl
U410010AM & F814W  & 1000 \nl\nl
\multicolumn{3}{c}{WFPC2 Jan. 10 1998} \nl
U4100201M & F336W  & 1900 \nl
U4100202M & F336W  & 600 \nl
U4100203M & F336W  & 1900 \nl
U4100204M & F336W  & 600 \nl
U4100205M & F555W  & 260 \nl
U4100206M & F555W  & 1000 \nl
U4100207M & F555W  & 1000 \nl
U4100208M & F814W  & 260 \nl
U4100209M & F814W  & 1000 \nl
U410020AM & F814W  & 1000 \nl\nl
\multicolumn{3}{c}{NICMOS/CAM2 Aug. 11 1997} \nl
N46D02APM & F222M  & 1279 \nl
N46D02B6M & F187N  & 2303 \nl
N46D02BHM & F190N  & 2303 \nl\nl
\multicolumn{3}{c}{NICMOS/CAM2 Sep. 9 1997} \nl
N46D03POQ & F160W  & 2559 \nl
N46D03PXQ & F160W  & 2559 \nl
N46D03Q6Q & F110W  & 2559 \nl\nl
\multicolumn{3}{c}{NICMOS/CAM1 Aug. 14 1998} \nl
N4WE01010 & F164N  & 2560  \nl
N4WE01040 & F166N  & 2560  \nl
N4WE01070 & F166N  & 2560  \nl
N4WE010A0 & F164N  & 2560  \nl\nl
\multicolumn{3}{c}{NICMOS/CAM3 Jun. 1998 (PID\tablenotemark{a}~7919) } \nl
N4K46KXJQ & F187N  & 768 \nl
N4K46KXKQ & F160W  & 192 \nl\nl
\multicolumn{3}{c}{NICMOS Archive (PID\tablenotemark{a}~7330) } \nl
N3ZB44010 & F160W  & 319 \nl
\enddata
\tablenotetext{a}{Proposal ID of GO Observations which we obtained
from the archive}
\end{deluxetable}
IR and CO observations of the dust lane can be modeled by a thin
warped disk (Quillen et al. \cite{quillen92};
Quillen, Graham \& Frogel \cite{quillen93}) which dominates ground-based 
near-IR observations along with the extended galaxy emission
(Packham et al. \cite{packham} and references therein).
R-band imaging polarimetry from HST with WFPC (Schreier et al.  
\cite{schreier96}; Paper I) are also consistent with dichroic 
polarization from such a disk. At large radii the radial light- 
profile is well fitted with a de~Vaucouleurs law, but until now the 
dust lane has prevented secure measurements of the profile within the 
central arcminute ($\sim 1$kpc) of the galaxy or of the nucleus 
itself.  Recent HST NICMOS observations in the K-band (Schreier et al. 
\cite{schreier98}, hereafter Paper II) have 
however revealed the presence of an unresolved (i.e. FWHM$< 0.2\arcsec$) 
source located at the nucleus of Centaurus A , superimposed on 
extended galaxy emission.  A $\sim$ 1\arcsec\ radius emission line 
region centered on the nucleus observed in \PA\ was interpreted as an 
extended accretion disk around the AGN.

In this paper, we report new HST observations of the central region of the
galaxy: two contiguous WFPC2 fields covering approximately 2\arcmin\
by 4\arcmin\ around
the nucleus in the F336W, F555W and F814W bands, and NICMOS Camera 1, 2,
and 3 observations of the nucleus and inner jet. We discuss the
implications of these new high spatial resolution data in U through K on
our understanding of the active nucleus, the jet, and the galaxy continuum
emission in the innermost regions of Centaurus A.

These observations are part of an extensive continuing program to 
study this giant elliptical galaxy -- the nucleus and jet of the 
nearest AGN, its stellar populations, and the relationship between its 
merger history and its nuclear activity.  The current paper 
concentrates on the photometric structure of the galaxy interior to the
dust lane, the spectral energy distributions of the nucleus, and the
search for an optical/ir counterpart to the radio jet.

In Section \ref{sec:observ} we discuss the observations and data 
reduction.  The observational results are described in Section 
\ref{sec:res}: Section \ref{sec:AGN} derives the location of the 
active nucleus and its spectral energy distribution; Section 
\ref{sec:feii} presents new $\FeII\lambda 1.64\mu$m observations of 
the \PA\ nuclear disk discovered in Paper II; Section \ref{sec:redden} 
presents a reddening map and reconstructs the inner galaxy light 
profile; Section \ref{sec:nic3} presents the Pa$\alpha$ map which 
identifies the presence of several star formation regions; and Section 
\ref{sec:knota} summarizes the observations related to the jet "Knot A".
Section \ref{sec:spnuc} discusses the implications of the 
spectral energy distribution of the nucleus.  Section 
\ref{sec:lightprof} analyses the galaxy light profile and its 
implications for the mass of the black hole.  Section \ref{sec:SBAGN} 
discusses the implications of these observations on the starburst-AGN 
connection, and on the black hole fueling mechanism.  Finally, 
section \ref{sec:jet} analyses the effect of the jet on the 
circumnuclear region.

\vskip 0.5cm
\section{Observations and Data Reduction}
\label{sec:observ}

Centaurus A (NGC 5128) was observed with both the Wide Field and Planetary
Camera 2 (WFPC2 -- Biretta et al. \cite{wfpc2}) and the Near Infrared
Camera and Multi-Object Spectrograph (NICMOS -- MacKenty et al.
\cite{nicmos}) on-board HST.  The WFPC2 data were taken on August 1, 1997 and
January 10, 1998; the two observations differed in telescope roll angle by
approximately 180\arcdeg\, optimizing coverage of the nucleus and dust lane
region in the WFPC2 field of view.  The pointings included the nucleus and
Knot A in the higher resolution PC chips.  The observations are logged in
Tab.~\ref{tab:logobs}.  Multiple exposures were performed in each of the
F336W (U), F555W (V) and F814W (I) filters, and final images were produced
by combining long and short exposures in each band, allowing removal
of cosmic rays and correction for saturated pixels in the longer exposures.

The data were recalibrated with the pipeline software {\it calwfp} in the
STSDAS/IRAF\footnote{IRAF is made available to the astronomical
community by the National Optical Astronomy Observatories,
which are operated by AURA, Inc., under contract with the U.S. National
Science Foundation. STSDAS contains HST-specific analysis routines
developed at STScI.} reduction package using up--to--date reference frames.
Warm pixels and cosmic rays  were removed (STSDAS tasks {\it warmpix} and
{\it crrej}) and mosaiced images with spatial sampling of 0\farcs1/pix
were obtained using the {\it wmosaic} task, which also corrects for the
optical distortions in the four WFPC2 chips.  The background was estimated from
areas with the lowest count rates in large, dust-extincted regions; flux
calibration
was performed using the conversion factors given by Whitmore (\cite{photflam}).

NICMOS Camera 2 (NIC2) observations of the nucleus of Centaurus A
were obtained on August 11, 1997 in the F222M (K), F187N (Pa$\alpha$)
and F190N (Pa$\alpha$ continuum) filters; these data
were extensively discussed in Paper II.  Deep NIC2 observations
of the X-ray/Radio feature ``Knot A'',
centered $\sim 20\arcsec$ NW of the nucleus, were obtained on
September 9, 1997 in the F160W (H) and F110W (J) filters.
NICMOS Camera 1 observations of the nucleus of Centaurus A
were obtained on August, 14 1998 in the narrow band filters
F164N and F166N (\FeII\ and adjacent continuum). NICMOS Camera 3
(NIC3) observations with the F160W and F187N filters were carried out
during the NIC3 campaign in June 1998 as part of a snapshot survey of
nearby galaxies (Proposal ID 7919).  Further NIC2 observations of the
nuclear region of Centaurus A with the F160W filter, obtained as part of a
GO snapshot survey (Proposal ID 7330), 
were retrieved from the HST Data Archive. All the NICMOS data were
recalibrated using the pipeline software CALNICA v3.1 (Bushouse et al.
\cite{calnica}) and the best reference files in the Hubble Data Archive as
of May 1998.  For further discussion of NICMOS data reduction see Paper II.
The NICMOS observations are logged in Tab. \ref{tab:logobs}.

To construct full field images, to compare features seen in different
observations taken at different times and with different instruments and,
especially, to create color maps, we calculated the
relative positions of the different pointings via cross correlation
techniques.  The WFPC2 observations taken at different dates were initially
aligned to one another using image WCSs (World Coordinate System, see Voit
et al. \cite{datahand} for a description) which provide the RA and Dec of
each pixel, based on the HST Guide Star Catalogue (GSC) reference system,
accurate up to $\pm$1\arcsec\ in absolute position.  Final alignment was
determined by fitting the positions of $\sim$10 stars in the overlap
region; the resulting images are found to be aligned to better than 0.3 WF
pixels (i.e. 0\farcs03).  The NICMOS and WFPC2 images were aligned again by
first using the image WCSs, and then cross-correlating selected regions with
well-defined morphologies.  The positions of stars observed in both WFPC2
and NICMOS images showed that the alignment was accurate to better than 0.4
NIC2 pixels ($\simeq 0\farcs03$).

\vskip 0.5cm
\section{ Results}
\label{sec:res}

\begin{figure*}
\caption{\label{fig:nucleus} Nuclear region of Centaurus A
in NICMOS (F222M, F160W) and WFPC2 (F814W, F555W) filters.
North is up and East is left.  The IR nucleus is located at the 0,0 position
and crosses are drawn to
help the eye in locating that position. A point source is detected
in the F814W and marginally detected in the F555W.  }
\end{figure*}
Figures \ref{fig:f814w}--\ref{fig:f336w} are gray-scale representations of
the WFPC2 mosaics in the filters F814W, F555W and F336W, corresponding to
the standard I, V and U bands. Figure \ref{fig:truecolor} is a true color
(Red=F814W, Green=F555W, Blue=F336W) representation of the above data.
The scale is 0\farcs1/pix.
\footnote{Color images available at http://www.arcetri.astro.it/~marconi}

The images reveal significant detail of the filamentary structure of the
dust lane as well as the presence of numerous sources over the entire field
of view. These comprise both unresolved point sources and many spatially
resolved globular clusters and stellar associations.  Particularly
prominent in the true-color image is a cluster of blue stars at the NW
upper edge of the dust lane.  Several clusters of blue stars are also
present within the dust lane and at its southern edge, clearly manifesting
active star formation (Dufour et al. \cite{dufour}, 
Paper I).  The prominent dark bands at the edges of
the dust lane presumably correspond to folds in the dusty molecular disk
proposed as responsible for the observed obscuration by Quillen, Graham \&
Frogel (\cite{quillen93}).
Photometry of the
stars and stellar associations in the data will be discussed in subsequent
papers.

\begin{figure*}
\centerline{
\psfig{file=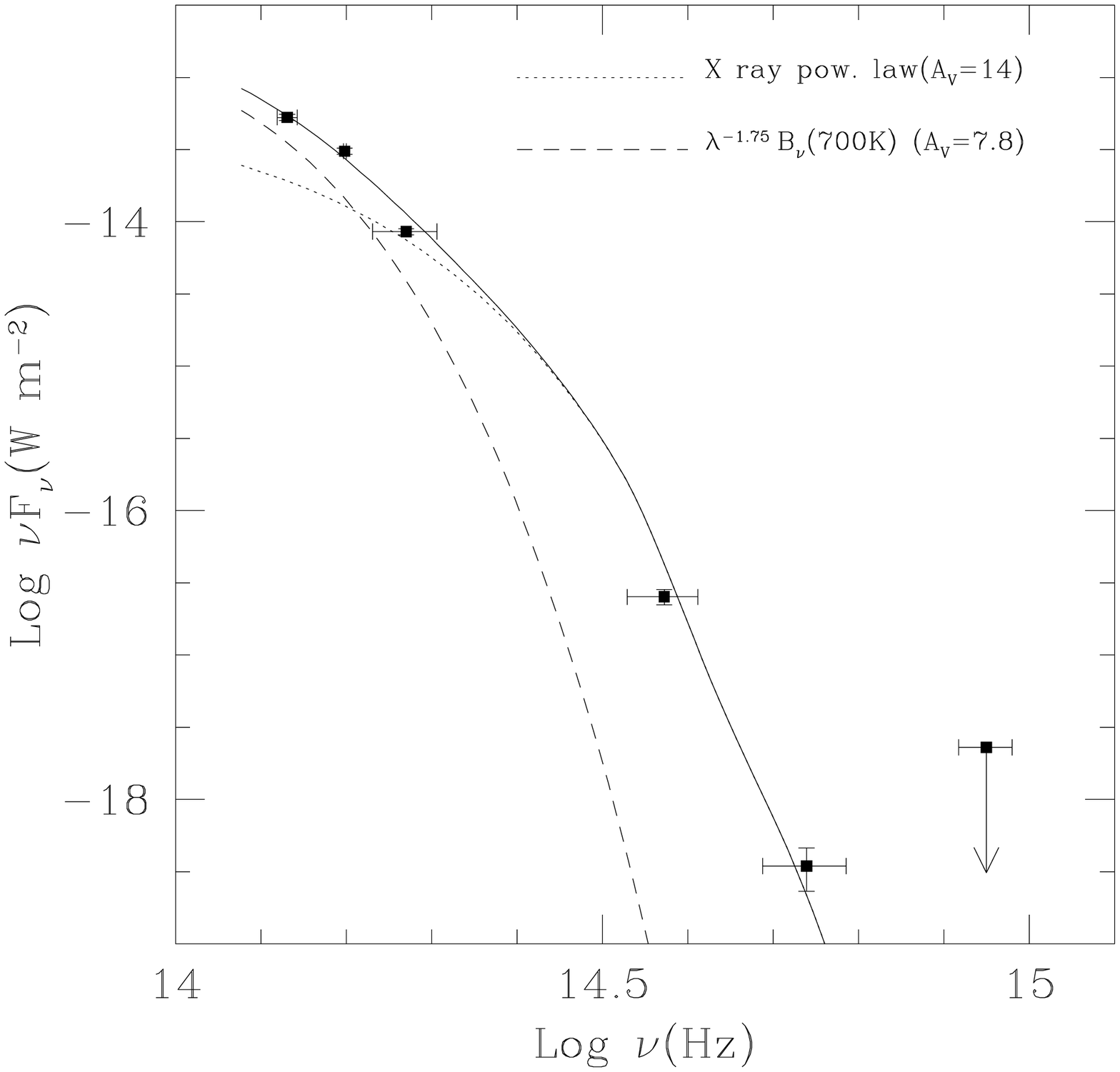,clip=yes,width=0.45\linewidth}
\psfig{file=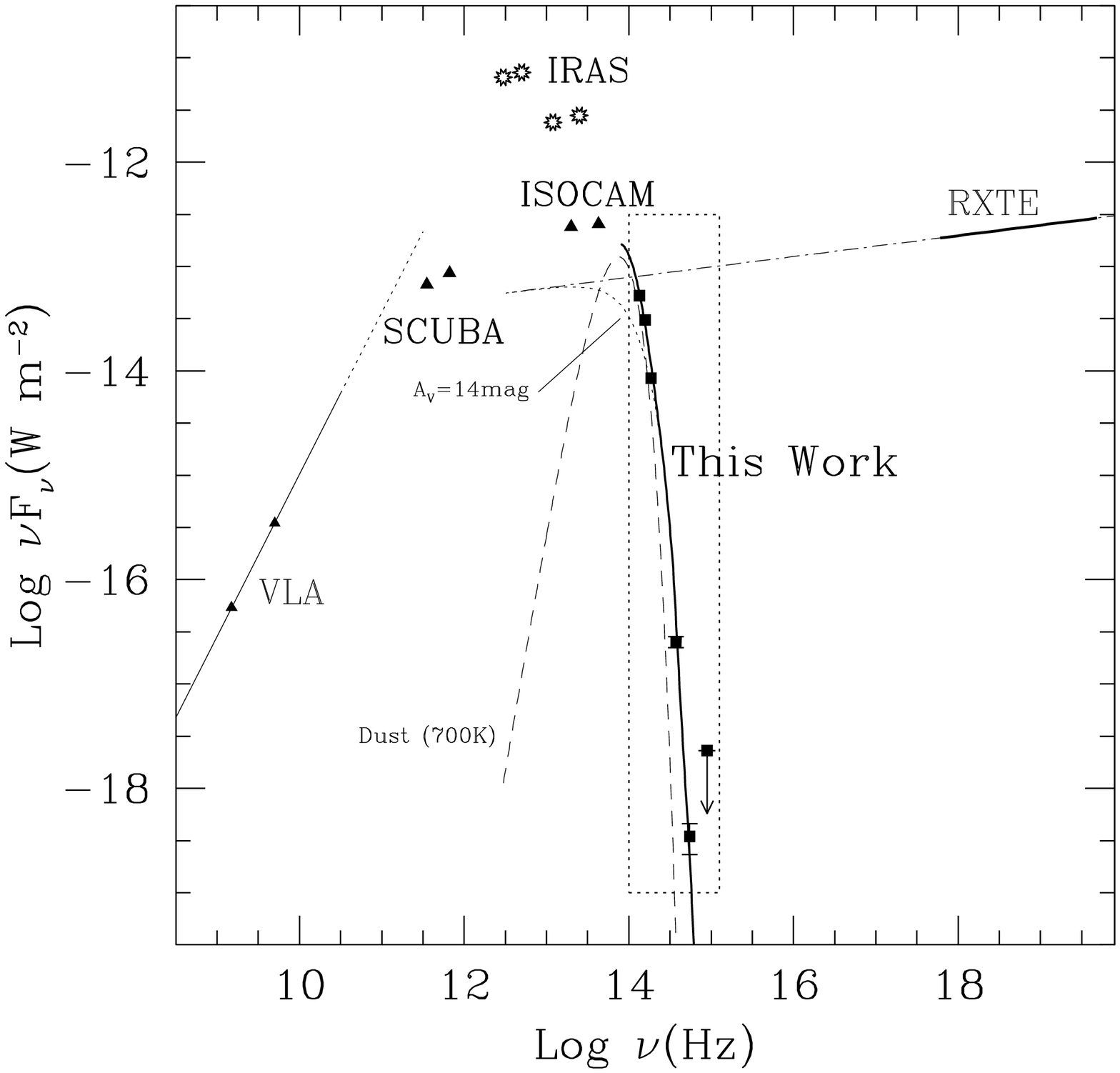,clip=yes,width=0.45\linewidth}}
\caption{\label{fig:spnuc}
Left Panel: Nuclear photometry with WFPC2/NICMOS.
Error bars along the x axis represent the width of the filters.
The solid line is the spectrum resulting from the combination
of (i) the power law extrapolated from the X-rays, reddened by $A_V=14$mag
(see also right panel),
and (ii) modified black-body emission at 700K (see text) reddened by
$A_V=7.8$mag.
Right panel: Overall nuclear spectrum of Centaurus A.
The thick line labeled ``RXTE'' represents the unabsorbed power law
observed in the 2.5--240 keV range which is then extrapolated at lower energies
(thin dash-dot-dash line).
The dotted rectangle marks the limits of the graph in the left panel
and the symbols in there (filled squares, thick, dashed 
and dotted lines) have the same meaning as above.
The stars are the IRAS photometric points while the filled triangles are the
SCUBA and ISOCAM photometric points from Mirabel et al. (1999). The
filled circles are the radio VLA points from Burn, Feigelson \& Schreier (1983)
and are connected by a sinchrotron spectrum. See also table 2 for 
values of the plotted photometric points.}
\end{figure*}
\vskip 0.5cm
\subsection{\label{sec:AGN} The location of the nucleus}

In Paper II, we identified a strong point source observed at 2.2$\mu$m as
the active galactic nucleus of Centaurus A, via its spatial coincidence with
the radio nucleus (within the $\pm1$\arcsec  absolute positional
uncertainty of the HST GSC). Although the current WFPC2 images in U, V and
I do not show a prominent galactic nucleus, even with the superb spatial
resolution of HST, the nucleus can be located in the optical data using the
accurately aligned IR and optical images (Sec. \ref{sec:observ}). Figure
\ref{fig:nucleus} shows the $7\arcsec\times 7\arcsec$ region centered
around the IR peak.  From left to right are shown: the NICMOS F222M data
with the prominent nucleus; the similarly prominent unresolved component in
the F160W image; the optical counterpart of the nucleus seen as a faint
point source in the WFPC2 F814W data; and a marginally detected source in
F555W.  To our knowledge, these WFPC2 data contain the first unambiguous
detection of the nucleus of Centaurus A at optical wavelengths.  The
nucleus is unresolved at all wavelengths, providing upper limits on the
FWHM of the nuclear emission region of $<0\farcs2$ (1.7pc) at 2.2$\mu m$
and $<0\farcs1$ (0.9pc) at 8000 \AA\ respectively.

The data allow us to derive the spectral energy distribution of the nucleus
from the optical to the near-IR where, at the high spatial resolution of
HST, the central source is easily isolated from the galaxy background.  The
SED is discussed further in Section \ref{sec:spnuc}
below.  Note that the extreme
steepness of the spectrum (the nucleus is $\sim 10{^5}$ times fainter in V
than in K) has an important impact on the accuracy of the nuclear broad
band photometry. The standard procedure used to flux--calibrate HST images
involves multiplying the count rate by the PHOTFLAM keyword (the conversion
factor to flux units in erg s$^{-1}$ cm$^{-2}$ \AA$^{-1}$).  However, this
conversion factor assumes a flat continuum (i.e. a constant $F_{\lambda}$)
which, clearly, does not apply to the Cen A nucleus, whose overall 0.5 --
2.2 $\mu$m spectrum can be approximated by a power law with spectral index
-6.6, i.e. $F_{\nu} \propto \nu^{-6.6}$.  Therefore, the flux scale must be
re-computed for each filter by convolving the filter transmission curve
with the source spectrum. We adopted an iterative procedure:
using the PHOTFLAM keywords we obtained zero order flux estimates, fitted a
reddened power law, and re--computed the PHOTFLAM keywords,  repeating
the process until it converged.  The decrease of the PHOTFLAM keyword 
was less than 1\% in F222M and F190N, but $\sim 7\%$ in F160W, and 40\% and
55\% for F814W and F555W respectively (see Table \ref{tab:spnuc}).  In
conclusion, it is essential to convolve the steep nucleus input spectrum
with the instrumental responses.  The resulting spectrum derived from our
corrected HST photometry is presented in Fig. \ref{fig:spnuc} (left panel)
and is discussed in Sec. \ref{sec:spnuc}.

\begin{figure*}
\caption{\label{fig:feiidisk} 
Surface Brightness contours of the NIC1, continuum subtracted \FeII\ image
overlayed on a grayscale representation of the NIC2 \PA\ image (Paper II).
The contours are logarithmically spaced by 0.2dex and the lowest
contour level is $1.5\times 10^{-15}$ erg cm$^{-2}$ s$^{-1}$ arcsec$^{-2}$.
The gray scales are between -2 and 25
in units of $10^{-15}$ erg cm$^{-2}$ s$^{-1}$ arcsec$^{-2}$.
North is up and East is left.  }
\end{figure*}
\vskip 0.5cm
\subsection{\label{sec:feii}The nuclear disk}

In Paper II, we presented NICMOS 2 observations in \PA\ of the nuclear
region of Centaurus A, which we interpreted as an inclined, $\sim 40$pc
diameter ionized disk. In Figure \ref{fig:feiidisk} 
we overlay on the \PA\ image the intensity contours
of continuum subtracted $\FeII\lambda 1.64\mu$m image.
Morphologically, the emission in \FeII\ and \PA\ are almost
indistinguishable and the \FeII/\PA\ ratio is $\sim 1/3$. 
The total detected \FeII\ emission of the disk 
(i.e. in a $0\farcs9\times2\arcsec$ aperture)
is $2.3\times 10^{-14}$ erg s$^{-1}$ cm$^{-2}$ which is
in excellent agreement with  the flux 
of $(2.5\pm0.4)\times 10^{-14}$ erg s$^{-1}$ cm$^{-2}$
in a $1\farcs4\times 1\farcs6$ aperture centered on the nucleus
found by Simpson \& Meadows (\cite{simpson}) with ground-based near-IR 
spectroscopy.

The ratio $\FeII/\PA=1/3$ corresponds to \FeII/Br$\gamma\sim 4$
(6.6 with $A_V=7$ and 12 with $A_V=15$). After taking into account extinction
the high \FeII/Br$\gamma$ value is typical of low excitation Seyfert
galaxies and LINERs (cf. Moorwood \& Oliva \cite{moorwood},
Alonso-Herrero et al. \cite{alonso}).
The much larger disks detected by HST at the centers of elliptical galaxies
usually have LINER spectra, a typical example being the gaseous disk
in M87 (e.g. Dopita et al. \cite{dopita}).

\vskip 0.5cm
\subsection{\label{sec:redden}
The extended continuum emission, the reddening correction, and
the galaxy light profile}

\begin{figure*}
\caption{\label{fig:reddening} F222M (K) image of the nuclear
region and E(B-V) map derived from the H-K colors.
Observed and dereddened F814W images (nuclear region only).
North is up and East is left. Each square has the size of the NICMOS 2\
field of view ($20\arcsec\times 20\arcsec$).  }
\end{figure*}
The X-ray spectral low energy cutoff has long provided strong evidence for
absorption along the line of sight to the Cen A nucleus.  It has also long
been clear from inspection of the optical images (cf. Fig.
\ref{fig:nucleus}) that the dust lane, interpreted as a warped disk of gas
and dust crossing the nuclear region, produces significant foreground
obscuration.  To study the SED of the nucleus, it is crucial to disentangle
the foreground (disk) component from the intrinsic local absorption near
the AGN.  The extinction due to the dust lane also prevents a direct
determination of the radial light profile near the nucleus of the galaxy in
the wavelength bands observed.

We can derive an average light profile only
by combining the available information in the F555W, F160W and F222M bands
after estimating the respective reddening corrections.
This is of course possible only assuming that no significant colour
gradients are present.

We briefly present a simple, zero-order estimate to correct for reddening
which, hypothesizing foreground screen extinction, is derived from the
color in two bands at $\lambda_1$ and $\lambda_2$ as:
\begin{equation}
M_{\lambda_1}^\circ = M_{\lambda_1}-A(\lambda_1) =
M_{\lambda_1}-\frac{R(\lambda_1)}{R(\lambda_1)-R(\lambda_2)}
E(\lambda_1-\lambda_2)
\end{equation}
where $M_{\lambda_1}^\circ$ and $M_{\lambda_1}$ are the intrinsic
and reddened magnitudes at the effective wavelength $\lambda_1$,
$R(\lambda)$ is the extinction curve, i.e. $A(\lambda)=R(\lambda) E(B-V)$,
and $E(\lambda_1-\lambda_2)=A(\lambda_1)-A(\lambda_2)$.
$E(\lambda_1-\lambda_2)$ can be derived from the color
$M_{\lambda_1}-M_{\lambda_2}$ as:
\begin{equation}
E(\lambda_1-\lambda_2) = (M_{\lambda_1}-M_{\lambda_2}) -
(M_{\lambda_1}^\circ-M_{\lambda_2}^\circ) 
\end{equation}
where we assume an intrinsic color,
$M_{\lambda_1}^\circ-M_{\lambda_2}^\circ$, constant over the entire field
of view. Different point spread functions (PSFs) are taken into account by
convolving each image with the PSF of the other. 
We have adopted the reddening curve by
Cardelli, Clayton and Mathis (\cite{cardelli})
according to which $R(\mathrm{F555W})=3.20$, $R(\mathrm{F814W})=1.88$,
$R(\mathrm{F160W})=0.58$ and $R(\mathrm{F222M})=0.35$.

\begin{figure*}
\centerline{
\psfig{file=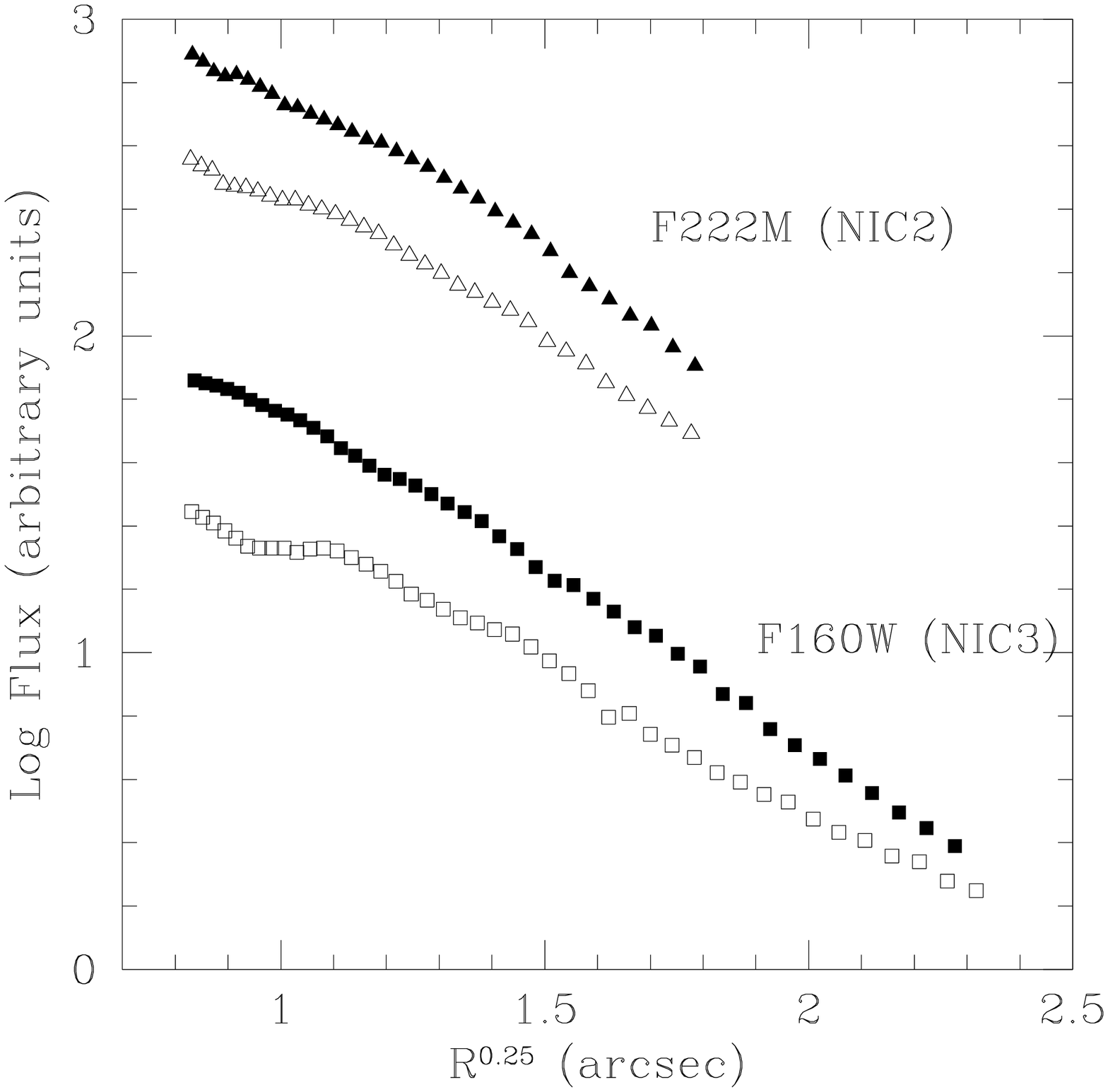,width=0.45\linewidth}
\psfig{file=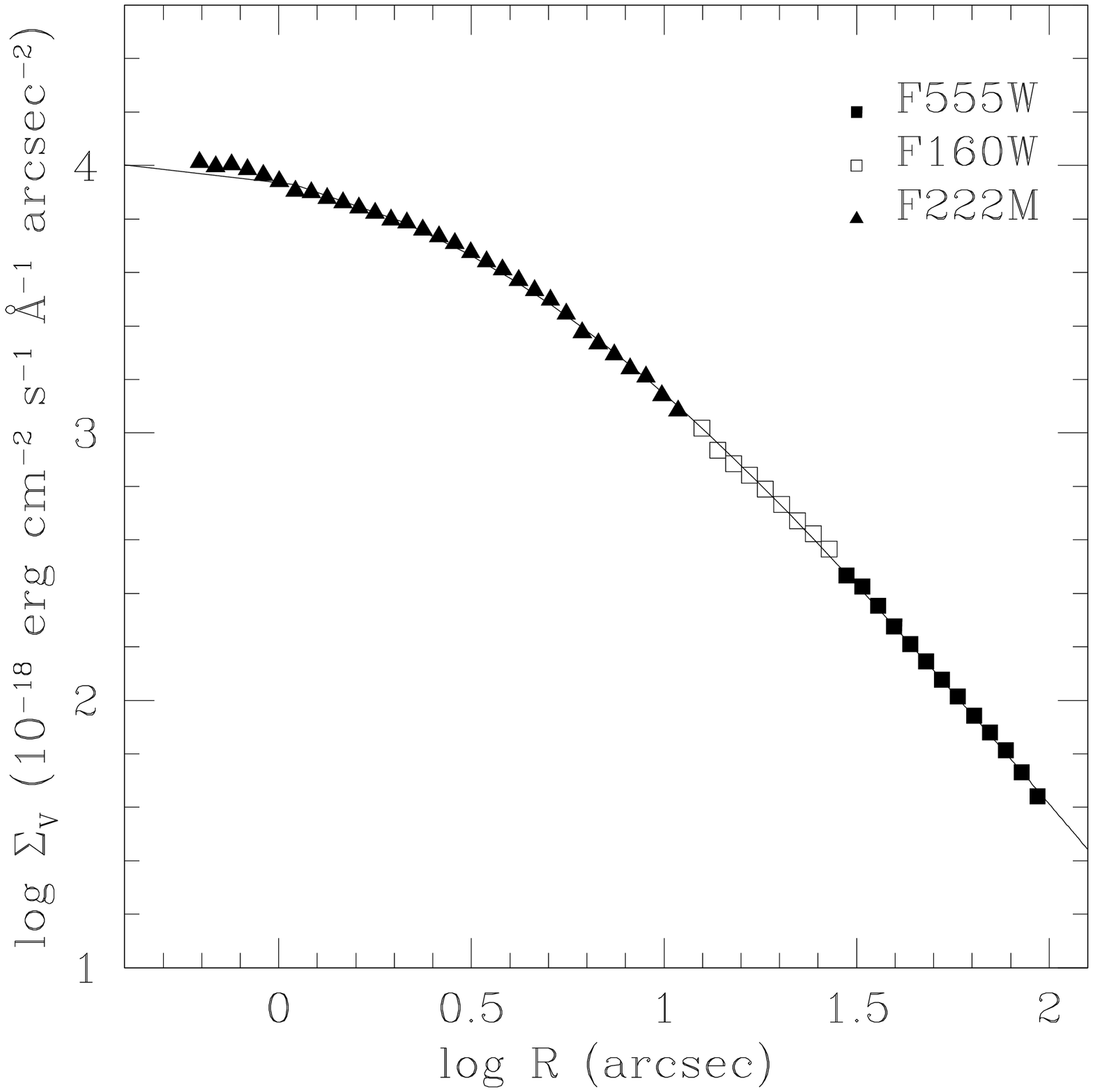,width=0.45\linewidth}
}
\caption{\label{fig:profiles}
Left Panel: Effect of reddening correction on the nuclear
light profiles for the F160w and F222M filter. Empty symbols
are the light profiles before reddening correction.
Right panel: Reconstructed nuclear profile in the V band.
Measurements from H and K band were rescaled to match the V band
observations. The solid line is the best fitting ``Nuker-Law'' (see text).}
\end{figure*}
Using the above formalism, we construct an E(B-V) map for the nuclear
region by combining the F160W (H) and F222M (K) NICMOS images. These are
more suitable for the reddening determination than the V and I band fluxes,
which drop to undetectable levels in many of the heavily reddened regions
of the dust lane. We assume an intrinsic colour $H-K=0.2$ which
is an average value for both spiral bulges and elliptical galaxies 
with a standard deviation of only 0.1 magnitudes
(e.g. Hunt et al. \cite{hunt} and references therein).
Note that the color correction due to the non-standard NICMOS filters
is negligible (0.2 in Jonhson's system corresponds to 0.24 in NICMOS).

To qualitatively verify the accuracy of the correction procedure, we apply
the reddening correction to the F814W WFPC2 image. Figure
\ref{fig:reddening} shows the F222M image, the E(B-V) image derived
from H-K, the uncorrected and reddening corrected
F814W images.  The corrected F814W image is seen to be
mostly smooth, similar to the F222M image which is obviously little
affected by reddening.  The correction fails only in a region of high
extinction below the nucleus (i.e. the fold of the molecular disk, see
Quillen, Graham \& Frogel \cite{quillen93}) where
E(B-V) reaches values higher than $\sim4$mag.  A small linear
structure starts below the nucleus and extends NW across the dust lane.
The extinction in the regions immediately surrounding the nucleus is
$E(B-V)\sim2.5$mag, in good agreement with the estimate of $A_V\sim 7$mag
given by Schreier et al. (\cite{schreier96}).
This value can be considered as a lower
limit to the extinction of the nuclear light, caused by foreground
extinction.  In the rest of the field, apart from patches like a knot north
of the nucleus ($E(B-V)\sim2$), the extinction generally decreases from
south to north, perpendicularly to the dust lane, reaching values as low
as 0.1.

The reddening corrections allow us to combine data taken in different bands
and reconstruct the galaxy radial light-profile.
We  combine the NIC2
(F222M) data, with its rather small ($r<10\arcsec$) field of view, with the
larger FOV NIC3 (F160W) data (reddening correction derived from the I-H
color) and the WFPC2 (F555W) data (reddening correction derived from the
V-I color).  Distortions caused by the patchy reddening extinction are well
corrected in the azimuthal average, as can be seen in Fig.
\ref{fig:profiles} (left panel), where the H and K band light profiles after
reddening correction (filled symbols) are smoother than those before
correction (empty symbols); as expected, the improvement is more evident in
the H band.

The overall radial profile of the nuclear region of is shown in Fig.
\ref{fig:profiles} (right panel).  The filled squares represent the points
derived from the F555W data, the filled triangles from the F222M data, and
the empty squares from the F160W data.  The light profiles from the F160W
and the F222M image are normalized to match that of the F555W image, using
scaling factors determined from fitting the light profiles.  We use the
``nuker-law,'' an empirical expression introduced by
Lauer et al. (\cite{lauer}), to
fit the core properties of elliptical galaxies observed at HST resolution:
\begin{equation}
I(r) = 2^\frac{\beta-\gamma}{\alpha}I_b\left(\frac{r_b}{r}\right)^\gamma
\left[ 1+\left(\frac{r}{r_b}\right)^\alpha
\right]^\frac{\gamma-\beta}{\alpha}
\end{equation}
where $I(r)\propto r^{-\gamma}$ at the center, 
$I(r)\propto r^{-\beta}$ at larger radii, and the transition occurs at
$r_b$ with surface brightness $I_b$. $\alpha$ represents the sharpness of
the transition.

The fit parameters, including the scaling factors from H to V ($S_H$) 
and K to V ($S_K$), are determined via $\chi^2$ minimization.
Since it is not possible to correctly normalize the $\chi^2$ in order
to perform a statistical analysis
(see sec. 4.1 of Carollo et al. \cite{carollo97}
for an extensive discussion) we estimated uncertainties
by mapping the $\chi^2$ in the parameter space. For a thousand times, 
we randomly selected the starting values of the fit parameters
and let the minimization algorithm converge to find all local
minima of $\chi^2$.
Acceptable fits were then selected by means of visual inspection.

The results of the 
fit are $I_b$=2.5--5 ($10^{-15}$ erg cm$^{-2}$ s$^{-1}$ \AA$^{-1}$ 
arcsec$^{-2}$), $R_b$=3.3\arcsec--5.6\arcsec, $\alpha$=0.8--1.6, 
$\beta$=1.6--2.0 and $\gamma$=0.0--0.3.  The best fit solid line in 
the figure has $I_b$=3.9, $R_b=3.9\arcsec$, $\alpha$=1.0, $\beta=1.8$, 
$\gamma=0.0$ and the scaling factors 
$\log$).  $S_H=0.50$ and $S_K=0.86$ (in $\log$).  The relatively large 
confidence bands for the model parameters arise because of 
the uncertainties in the scaling factors and the small range of radius 
available for fitting.  Never-the-less, the profile appears to fit the 
data well.  We discuss the implications of the radial profile in 
Section \ref{sec:lightprof}.

\begin{figure*}
\caption{\label{fig:nicmos3}
NICMOS 3 F160W (Left Panel) and continuum subtracted
Pa$\alpha$ (right panel) images.
North is up and East is left. Each square has the size of NICMOS 3
field of view ($50\arcsec\times 50\arcsec$).
VLA radio contours are overlayed on the Pa$\alpha$ image (see text).  }
\end{figure*}
\vskip 0.5cm
\subsection{\label{sec:nic3} Pa$\alpha$ emission in the dust lane}

The NIC3 F160W image and a continuum subtracted Pa$\alpha$
(F187N) image are presented in Figure \ref{fig:nicmos3}, along with a 6 cm
radio contour overlay (discussed further in the next section).  Since no
NIC3 narrow band continuum image is available, we derive one from the F160W
image. To account for reddening, we estimate E(B-V) by comparing the F160W
image with the WFPC2 F814W image (the F222M image used earlier has too
small a field of view) and then apply the differential reddening correction
to F160W to match that of F187N.  The results, as seen in the figure,
appear reasonable -- most of the field is devoid of emission, and the
structures due to reddening which appear in a direct line-continuum
subtraction have disappeared.  Moreover, the morphology of the nuclear
region, with the disk and bow-shock like features, is the same as seen in
the NIC2 Pa$\alpha$ continuum-subtracted data, as presented in Paper II.

Two chains of emission line knots are observed above and below the 
nucleus, roughly parallel to the dust lane.  These are located in the 
regions of higher extinction -- on the folds of the putative warped 
disk.  Apart from the Pa$\alpha$ emission knots close to the nucleus, 
already extensively discussed in Paper II, the radio overlay shows no 
obvious correlation with the jet, in contrast with previous HST 
observations of Seyfert or radio galaxies with radio jets
(e.g.  Capetti et al. \cite{capetti}, Axon et al. \cite{axon98}
and references therein).
The fluxes from the more intense Pa$\alpha$ 
knots are of order $\sim 2\times 10^{-14}$ erg cm$^{-2}$ s$^{-1}$ in a 
$2\arcsec\times2\arcsec$ area, corresponding to $\sim 5\times10^{50}$ 
ph s$^{-1}$ H-ionizing photons (using D=3.5 Mpc and A$_V=7$mag as a 
reddening estimate in the regions of star formation).  The 
total flux associated to these knots is $\sim 1.3\times 10^{-12}$ erg 
cm$^{-2}$ s$^{-1}$.

\vskip 0.5cm
\subsection{\label{sec:knota} The Jet and "Knot A"}

There is a remarkable correspondence between the radio jet and the
broad-band V-I color map, as seen in Fig. \ref{fig:VIradio}.  A relatively
blue linear feature coincides with the trajectory of the radio jet from the
nucleus to the so-called ``Knot A'', well known since early X-ray/radio 
observations and stretching from 15\arcsec\ to 30\arcsec\ NW of the nucleus,
as indicated in the figure.
Faint emission in this ``blue channel'' is also apparent
in the F336W image (upper right panel in Fig. \ref{fig:VIradio}).

The F160W and F110W NIC2 images of the region around Knot A are displayed
in Fig. \ref{fig:knota} along with the F814W and F555W WFPC2 images
(rescaled and aligned via the cross-correlation technique described in Sec.
\ref{sec:observ}).  The figure also overlays 6cm VLA radio contours (E.
Feigelson, private communication) on the F110W-F160W and F555W-F814W
images.  The radio is overlayed on the optical--infrared image by aligning
the radio and infrared nuclei. Although there is no obvious optical
emission correlated with the radio knots, there does appear to be a
correlation between the radio and the near--IR/optical colors: the radio
contours lie in a region of bluer continuum emission, similar to that
observed in the linear region discussed above.  Particularly intriguing is
the presence of redder spots, apparently coincident with bends in the radio
jet. These red spots might represent dense, dusty clouds which deflect the
radio jet. It is worth noting that we do not detect any line or continuum
emission associated with this putative interaction region.

We have investigated the possibility that continuum emission from
the jet might be present but "masked" by galaxy emission and the patchy
extinction.  Using as a template the spectrum of a blue, bright region far
from Knot A, we find that within statistical errors, knot A has the same
spectral shape, but with different scaling factors and E(B-V), implying no
significant emission from the jet itself. Note also that an extrapolation
of the radio synchrotron spectrum to the optical would not predict any
detectable flux.

\vskip 0.5cm
\section{\label{sec:discuss} Discussion}

\begin{deluxetable}{lcccr}
\tablewidth{0pt}
\tablecaption{\label{tab:spnuc} Nuclear spectrum of Centaurus A.}
\tablehead{
\colhead{Instrument} & \colhead{Wavelength} &
\colhead{Flux} & \colhead{$\Delta F/F$} & \colhead{Notes}}
\footnotesize
\startdata
\multicolumn{5}{c}{This Work\tablenotemark{a}} \\
NICMOS/F222M & 2.218    & 2.38E-15      & 5\%   & 0.99$\times$PHOTFLAM \\
NICMOS/F190N & 1.900    & 1.62E-15      & 5\%   & 1.00$\times$PHOTFLAM \\
NICMOS/F160W & 1.610    & 5.31E-16      & 5\%   & 0.93$\times$PHOTFLAM \\
WFPC2/F814W  & 0.803    & 3.15E-18      & 12\%  & 0.60$\times$PHOTFLAM \\
WFPC2/F555W  & 0.547    & 6.34E-20      & 33\%  & 0.45$\times$PHOTFLAM \\
WFPC2/F336W  & 0.337    & $<$6.8E-19    & --    & 1.00$\times$PHOTFLAM \\
\multicolumn{5}{c}{Mirabel et al. 1998\tablenotemark{b}} \\
ISOCAM/LW2   & 7        & 0.6           &       &               \\
ISOCAM/LW3   & 15       & 1.2           &       &               \\
SCUBA/450    & 450      & 13            &       &               \\
SCUBA/850    & 850      & 19            &       &               \\
\multicolumn{5}{c}{IRAS photometry
compiled in Mirabel et al. 1998\tablenotemark{b}} \\
IRAS         & 12       &      11.2     &   \\
IRAS         & 25       &      20.1     &   \\
IRAS         & 60       &      145      &   \\
IRAS         & 100      &      217      &   \\
\multicolumn{5}{c}{Rothschild et al. 1998\tablenotemark{c}} \\
RXTE/PCA/HEXTE & 2--10  & 3.40E-10    & 2\%   & Phot. Index $\Gamma\simeq1.9$ \\
\multicolumn{5}{c}{Burns, Feigelson \& Schreier 1983\tablenotemark{d}} \\
VLA            & 20     & 3584 & 3\% \\
VLA            & 6      & 6984 & 3\% \\
\tablenotetext{a}{Wavelength in $\mu m$; Flux in erg cm$^{-2}$ s$^{-1}$
\AA$^{-1}$; Notes are the conversion factor used in units of PHOTFLAM}
\tablenotetext{b}{Wavelength in $\mu m$; Flux in Jansky}
\tablenotetext{c}{Range of Wavelength in keV; Flux in erg cm$^{-2}$
s$^{-1}$}
\tablenotetext{b}{Wavelength in cm; Flux in milli-Jansky}
\enddata
\end{deluxetable}
\vskip 0.5cm
\subsection{\label{sec:spnuc}
The Active Nucleus and its Spectral Energy Distribution}

Fig. \ref{fig:spnuc} (right panel) compares our photometry of
the nucleus of Centaurus A with selected measurements from 
the literature (see also Table \ref{tab:spnuc}).

In X-rays, the 2.5--240 keV spectrum observed by RXTE is
well fitted by a power law (plus a Fe K--shell line)
with photon index $\sim$1.9, absorbed by a column density of
$9.4\times 10^{22}$ cm$^{-2}$ (Rothschild et al. \cite{rxte}).
In the mid-IR submm range, we consider the recent ISO and
SCUBA observations of Mirabel et al. (\cite{iso}).
We also plot the fluxes reported by IRAS although, with its
relatively poor angular resolution, these points are dominated by the 
extended galaxy emission;
as shown by ISO, the nucleus contributes at most 10\% of the total 
mid-IR flux (Mirabel et al. \cite{iso}).
Indeed, the ISO points themselves overestimate the contribution from 
nucleus, since even the smallest 7\arcsec\ aperture includes significant
extended galaxy emission.
At radio wavelengths we employ the VLA fluxes of the nucleus of Burns, Feigelson
and Schreier (\cite{burns}).

The absorption toward the nucleus of Centaurus A
is inferred both from X-rays and infrared observations.
In X-rays the observed absorbing column density
corresponds to a visual extinction of $A_V\sim 47$ mag
(assuming a standard gas-to-dust ratio $A_V=5\times 10^{-22} N_H$).
We estimate $A_V\sim 15$ mag applying the extinction curve
of Draine and Lee (\cite{draine}) to the 
the 10$\mu m$ silicate absorption feature in the ISO spectrum
(Mirabel et al. \cite{iso}).
The two values closely follow the relationship observed in
Seyfert galaxies which, on average, have $A_V^{IR}\sim (0.1-0.5)\, A_V^{X}$,
where
$A_V^{IR}$ is the extinction determined from IR spectra and $A_V^{X}$
is determined from X-ray absorbing column densities, with the assumption of
galactic dust-to-gas ratio (see Sec. 3 of Granato, Danese \& Franceschini
\cite{granato}).
We note that in Section \ref{sec:redden} above we have estimated a lower
limit for the extinction toward the nucleus of $E(B-V)>2.5$.

The presence of a non-thermal source in Cen A is well established from
both the X-ray and radio observations and for this reason, a power law fit
of the SED of the nucleus is a viable starting hypothesis.  Indeed, Packham
et al. (\cite{packham}) described the IR nuclear spectrum of Cen A as a
power law reddened by 16 mag.
However, an extrapolation of their fit underestimates
our V band flux by more than a factor of 1000 while, at the same time,
over-predicting the 10 $\mu$m ISO flux by a factor of 20. This drawback is
common to any reddened power law -- reddening can not reproduce the sharp
break at $\lambda < 3.5 \mu$m unless the absorption is as large as
E(B-V) $\sim 20$, making the optical emission unobservable.
An intrinsically steep
spectrum is also unacceptable since the observed spectrum at long
wavelengths, where reddening is negligible, is essentially flat.

On the other hand, an extrapolation to the optical
of the observed X-ray power law,
reddened by $A_V=14$mag (consistent with the extinction derived from the
silicate feature and from the X-ray extinction), accounts for the V and I
measurements without exceeding the IR points (dotted line
in Fig. \ref{fig:spnuc}) which, therefore, must be accounted for
by another component.

The infrared emission of AGNs at wavelengths longer than 2$\mu$m is usually
interpreted as thermal emission from hot dust close to the active nucleus
(e.g., Rieke \& Lebofsky \cite{rieke}).
The observed SED of the Cen A nucleus in
the mid-infrared is also qualitatively consistent with the presence of warm
dust, while the sharp drop of the nuclear intensity longward of 3 $\mu$m is
reminiscent of dips seen in quasar spectra, usually ascribed to the cutoff
in the dust temperature distribution due to the sublimation of grains
(Sanders et al. \cite{sanders}).
We note that silicates sublime at $T \sim$ 1400 K
and graphite at $T \sim$ 1750 K.  We test quantitatively if dust emission is
consistent with the observed SED by estimating the dust temperature,
$T_{dust}$, which corresponds to the colors measured in adjacent bands:
\begin{equation}
\frac{F_{\nu_1}}{F_{\nu_2}} = \frac{\lambda_1^{-\beta}}{\lambda_2^{-\beta}}
\frac{B_{\nu_1}(T_{dust})}{B_{\nu_2}(T_{dust})}
\end{equation}
where $F_{\nu}$ is the dereddened flux, $B_\nu(T_{dust})$ is
the blackbody function and the dust emissivity is
$\epsilon_\nu\propto \nu^{-\beta}$.  $\beta$ is usually in the
range 1.5-2 and here we assume $\beta=1.75$.
The derived temperature increases from $T \sim 250$ obtained from 
$F(7\mu m)/F(15\mu m)$, to $T \sim 920$K for I-H, and to $T \sim 1300$ for
V--I.  This result is consistent with the idea that we are seeing dust with
a broad temperature distribution, and that at shorter wavelengths we are
sampling regions of higher temperature.
However, as we have shown above,
the nucleus of Cen A suffers a foreground extinction of at least
E(B-V) $\sim 2.5$ and the reddening corrected 
V-I slope becomes incompatible with thermal emission 
since the corresponding temperature $T \sim 2800$ exceeds
the dust sublimation temperature.

The spectrum drawn with a thick solid line in Fig. \ref{fig:spnuc}
(left and right panel) is the combination of two components:
the extrapolation of the X-ray power law reddened by $A_V=14$ mag and
dust emission at 700K reddened by $A_V=7.8$ mag, i.e. $E(B-V)=2.5$ the
minimum foreground extinction to the nucleus.
Dust emission is described by $\lambda^{-1.75}B_\nu(700\mathrm{K})$.
This composite spectrum well reproduces the observed points: 
the emission in V and I is completely "non-thermal", 
that in K is mostly
($\sim 70\%$) thermal, while in the H band the two emission processes
are comparable. Of course these results are only indicative and the emission
in the near infrared could be completely dust dominated.
Note that even if the 700K component fits the observed HST data points, lower
temperature components are required in order to fit the ISO and SCUBA
measurements. However, a complete analysis of the infrared emission spectrum
is beyond the aims of this paper.

Therefore our data points on the nucleus are consistent with a scenario
in which dust dominates emission in the near infrared
but non-thermal emission accounts for the V and I bands and
contributes to the near-IR emission.
This is in agreement with previous suggestions
that Centaurus A hosts a misaligned BL LAC nucleus
(e.g. Bailey et al. \cite{bailey}, Morganti et al. \cite{morganti}, 
Steinle et al. \cite{steinle} and references therein).
Near-IR polarimetric observations at HST resolution can further test
this scenario.

Turner et al. (\cite{turner}) reported that the nucleus of
Centaurus A at 3.5 $\mu$m decreased its flux by a factor 2.5 in less than 5
years from 1987 to 1992 (comparing their data with those of L\'epine et al.
\cite{lepine}).
Comparing the surface brightness profiles and performing aperture
photometry in our F160W and F187N images, obtained over 9 and 10 month
baselines, respectively, we find no evidence for variability of the
nucleus above the 10\% level.

We note that even at the highest spatial resolution --
0\farcs1, corresponding to 1.7pc or $\sim 5$ years --
the variability reported by
Turner et al. (\cite{turner}) 
is not necessarily indicative of non-thermal emission,
since the timescales are also consistent with variability in dust
emission. Using the formula by Barvainis (\cite{barvainis})
the dust sublimation radius
is $\simeq 0.05$pc (with $L_{AGN}=10^{10} \mathrm{L}_\odot$)
therefore only variability on time scales shorter than $\sim 2$
months can unambiguously
prove the presence of a non thermal component in the near-IR.

For completeness, even if the proposed scenario accounts for many of
the properties of the Centaurus A nucleus,
we note that emission in the V band could also be accounted for 
by a central star cluster.

\begin{figure*}
\caption{\label{fig:VIradio}
Overlay of the VLA radio observations on V--I grayscales.
The field of view in both panels is
$\sim 40\arcsec\times 35\arcsec$.
Note the correspondence between a ``blue channel'' and the radio contours.
Knot A indicates a well known X-ray/radio feature stretching 
from 15\arcsec\ to 30\arcsec\ NW of the nucleus.
The panel at the upper right corner represents the U grayscales where
emission with a linear morphology aligned with the radio jet is
marginally detected.
}
\end{figure*}
\vskip 0.5cm
\subsection{\label{sec:lightprof}
The galaxy light profile and the mass of the central Black Hole}

According to the classification by Faber et al. (\cite{faber}),
the brightness distribution of Centaurus A derived earlier is consistent 
with a ``core''
profile, with $\gamma\le 0.3$, and with a break at a radius of $\sim 3.9$.
In general, galaxies with core profiles have $M_V<-22$, while those with
power law profiles have $M_V>-20.5$, and both types can exist at
intermediate values.  Fitting a de Vaucouleurs law to the V-band isophotes
of the galaxy outside the dust lane, Dufour et al. (\cite{dufour}) found
$R_e(V)= 305\arcsec$ and
estimated the total V magnitude for the elliptical component of Centaurus A
as 6.20 mag which, when corrected for E(B-V)=0.1 of foreground extinction,
becomes V$\sim 5.9$ mag.  From our profile fitting of the inner region, we
estimate the contribution of the central cusp, and find that the total flux
is increased by $\sim 15\%$ with respect to a simple de Vaucouleurs
law extrapolated in to the nucleus.  This corresponds to -0.15 in magnitude
i.e. V$\sim 5.75$ mag.  Using a Cen A distance modulus of
$(m-M)_\circ=27.7$ (Hui et al. \cite{hui}), we find $M_V=-21.95$ (i.e. log L
= 10.8 in solar units), consistent with Faber's core profile classification.

Van der Marel (\cite{marel}) has recently shown that the Faber classification is
consistent with the hypothesis that (i) all early-type galaxies have
central BHs that grew adiabatically in homogeneous isothermal cores and
(ii) these ``progenitor'' cores follow scaling relations similar to those
of the fundamental plane. This model suggests a roughly 
linear correlation between the
mass of the central black hole and the V-band galaxy luminosity $\log
M_{BH} \approx -1.83 + \log L_V$ in solar units (with an RMS scatter of
0.33 dex, or a factor two in the black hole mass).
This relationship is in agreement with the average relationship for nearby galaxies
with kinematically determined black hole masses
(e.g. Magorrian et al. \cite{magorrian}). Applying the above
hypotheses to the case of Centaurus A, we estimate the black hole mass to
be  $10^9$  M$_\odot$ with a factor $\sim$2 RMS uncertainty.  Given the
large uncertainties, this value is well in agreement with a dynamical mass
of $4.4\times 10^8$ M$\odot$ within 40pc, derived from H$_2$ spectroscopy
(Israel \cite{israel}).

\begin{figure*}
\caption{\label{fig:knota}
Upper Panels: from left to right, H, J and J--H NICMOS 2 images of the
Knot A region centered $\sim 20\arcsec$ NW of the nucleus.
The field of view is $\sim 20\arcsec\times 20\arcsec$.
VLA radio contours are overlayed on the J--H color image and show the 
stronger structures constituting Knot A.
Lower Panels: from left to right, I, V and V--I WFPC 2 images. As above,
radio contours are overlayed on the color image.
}
\end{figure*}
\vskip 0.5cm
\subsection{\label{sec:SBAGN}
The Starburst-AGN connection and the fueling of the AGN}

The Pa$\alpha$ emission we observe in the dust lane strongly suggests
that active star formation is taking place in the circumnuclear region.
We can estimate the star formation rate from \PA\ in the inner 
$50\arcsec\times 50\arcsec$ (i.e. NIC 3 field of view).
As shown in section \ref{sec:nic3},
the total observed \PA\ flux is $1.3\times 10^{-12}$ erg cm$^{-2}$ s$^{-1}$,
which, assuming a mean extinction of $A_V=7$, corresponds to an ionizing
photon rate of $Q(H)\sim3.1\times 10^{52}$ ph s$^{-1}$.
According to the models by Leitherer \& Heckman (\cite{leitherer}),
the inferred Star Formation Rate (SFR) depends mostly
on the assumed IMF (Initial Mass Function) slope and upper mass cutoff
and just slightly on metallicity.
With a Salpeter IMF and 30M$_\odot$ upper mass cutoff, the continuous SFR
is 0.3M$_\odot$ yr$^{-1}$. This further decrease to 0.1M$_\odot$ yr$^{-1}$
if the upper mass cutoff is 100M$_\odot$, but increases to 1M$_\odot$ yr$^{-1}$
with the same upper mass cutoff and a steeper IMF.
In any case we find a moderately high SFR in a small ($R<450$pc) 
region.

Star formation in the nuclear environment of AGNs is not uncommon.
In particular local Seyfert 2 galaxies, i.e. obscured AGNs, appear
characterized by enhanced circumnuclear star formation.
Observational evidence comes from the strength of the far-IR continuum
(Rodriguez-Espinoza et al. \cite{rodriguez}),
from mid-IR 10$\mu$m data (Maiolino et al. \cite{maiolino95}),
from the nuclear
mass-to-light ratio (Oliva et al. \cite{oliva}),
from narrow band optical images 
(Gonzales-Delgado et al. \cite{gonzales97})
and from the detection of UV spectral
features typical of O stars in a sample of Seyfert 2 galaxies
(Heckman et al. \cite{heckman}, Gonzales-Delgado et al. \cite{gonzales98}).

A likely explanation of this Starburst-AGN connection
is that, in order to feed the AGN, large amounts of gas must
be dumped in the inner Kpc of an active galaxy.
The processes which can accomplish this (galaxy interactions, bars etc.,
see below) also trigger star formation in the dense, concentrated gas.
A spectacular example of this process is given by
Maiolino et al. (\cite{maiolino99}) who
recently discovered a nuclear gaseous bar in the Circinus galaxy,
a well known Seyfert 2.

The observed detailed morphology in \PA\
suggests two parallel chains of star formation
located in regions of high extinction N and S of the nucleus.  We can
relate this morphology both to the warping of a molecular disk comprising
the dust lane, and also to the bar-like structure recently observed by ISO.

Quillen et al. (\cite{quillen92}) suggested that the regions of high
extinction above and below the nucleus are folds in the thin, warped
molecular disk evidenced by CO emission. The extinction is higher at the
folds because the disk is nearly edge-on. If star formation is taking place
in this molecular disk, then the Pa$\alpha$ morphology is also easily
explained, since a higher number of HII regions are projected along our
line of sight.

However, the Pa$\alpha$ emission is also seen to be
correlated with the edges of the putative bar observed with ISO at 7 and 15 $\mu$m
(Mirabel et al. \cite{iso}), as shown in Figure \ref{fig:isopa}.  This
correspondence is not unexpected, since the 7$\mu$m emission is presumably
dominated by PAH re-radiation, which could be excited by the hot young
stars seen in Pa$\alpha$.  Note that the more diffuse IR emission can be
attributed to older (B or A) stars which still emit enough UV photons to
excite the PAHs, but not enough to create the giant HII regions observed. 

If the IR morphology is indeed indicative of a bar, as suggested by Mirabel
et al. (\cite{iso}), the observed morphology is consistent with the overall
scenario which relates the onset of activity in galactic nuclei to
interactions between galaxies (eg., Baade \& Minkowski \cite{baade};
Toomre \& Toomre \cite{toomre};
Heckman et al. \cite{heckman86}; see also Balick \& Heckman \cite{balick}, and
Barnes \& Hernquist \cite{barnes} for reviews).
Specifically, ``minor mergers''
between a host galaxy and a small gas-rich companion or dwarf galaxy are
regarded as an important means of fueling active nuclei (eg., Mihos \&
Hernquist \cite{mihos};
Hernquist \& Mihos \cite{hernquist}; Walker et al. \cite{walker};
Bahcall et al. \cite{bahcall};
De Robertis, Yee \& Hayhoe \cite{derobertis}), in which case $\sim
10^8$~$M_{\odot}$ of stars, gas and dust can be dumped in toward the
central regions, providing a fuel supply for the AGN.  Such a scenario has
long been invoked for Cen A (cf. discussion in Paper II).

Galaxy interactions are also known to be appealing mechanisms for
triggering star formation via shocks induced in molecular clouds.  Detailed
n-body/hydrodynamic  models suggest that interactions between ellipticals
and small gas-rich galaxies result in the accretion of gas into the central
regions of the elliptical (eg. Weil \& Hernquist \cite{weil}; 
Mihos \& Hernquist \cite{mihos96}),
and that star formation can then be triggered via accretion shocks
in molecular clouds.  The gas is driven toward the center of the galaxy in
response to tidal forces (Binney \& Tremaine \cite{binney};
Barnes \& Hernquist \cite{barnes}).
In Centaurus A, we see star formation regions both within the dust lane
and at the edges of the putative bar.

Models predict shocks predominantly at the leading edges of such bars (cf.
Athanassoula \cite{athanas92a}, \cite{athanas92b}).
The shocks are likely to trigger star
formation, explaining the observed morphology, wherein the HII region are
located at the edges of the bar and along the "spiral arms".  At smaller
scales, the Pa$\alpha$ "disk" around the nucleus, discussed in Paper II,
could then be supplied by the gas streaming in from the bar, ionized by the
AGN.  Note that the orientation of the Pa$\alpha$ feature is perpendicular
to the edges of the bar.

In summary, the feeding of the AGN requires gas to be driven in toward the
nucleus, and the same galaxy interaction/merger which enriches the
elliptical galaxy with gas, can trigger star formation via shocks induced
in molecular clouds.

\begin{figure*}
\caption{\label{fig:isopa} Left Panel: Contours from the ISOCAM image
at 7$\mu$m. Center Panel: overlay of the ISOCAM contours on the 
NIC3 \PA\ image showing the morphological association between the
\PA\ emission and the edges of the putative bar. The small right panel
is the \PA\ disk from Paper II. Note that its major axis
is perpendicular to the edges of the ``bar''.  }
\end{figure*}
\vskip 0.5cm
\subsection{\label{sec:jet}
The Jet and the sweeping of the Interstellar Medium}

The data presented in Sec. \ref{sec:knota} provide no evidence for synchrotron
emission from the jet, either in the channel leading to Knot A or
from Knot A itself.  We note that an extrapolation of a simple synchrotron
spectrum from the radio would not produce detectable optical synchrotron
emission (see e.g. Paper I).
In this regard, note also that we
see no Pa$\alpha$ related to the jet, except possibly within a few
arcseconds of the nucleus (see Paper II).  We thus exclude emission from
young stars whose formation might have been triggered by the interaction
between the jet and the ISM.
We also note that we cannot exclude free-free emission from hot shocked gas.

We suggest that the ``blue channel''
corresponds to a region of gas ``evacuated'' by the jet, causing lower
extinction than the surroundings.
We can investigate the feasibility of this hypothesis -- the evacuation by
the jet of a region of gas corresponding to the observed extent of the blue
feature -- given what is known about the lifetime of the source and the
properties of the jet. The detailed
radio studies by Schreier, Burns \& Feigelson (\cite{schreier81}),
Burns, Feigelson \& Schreier (\cite{burns}),
Clarke, Burns \& Feigelson (\cite{clarke86}) and Clarke, Burns \&
Norman (\cite{clarke92}),
suggest that the bulk kinetic power of the northern jet is
$L_j \sim 1.8 \times 10^{43}$~erg~s$^{-1}$, while the minimum age estimate
of the inner lobes is $\tau_{\rm rad} \sim 5 - 10 \times 10^6$~yr. A
simple, order-of-magnitude estimate of the amount of energy required to
evacuate material from this region assumes it is described by a cylindrical
geometry with a diameter $\sim 5\arcsec$ ($\sim 85$pc) and length $\sim
30\arcsec $ ($\sim 500$~pc), initially filled uniformly by an ISM with
density $n_{\rm ISM} \sim 10$~cm$^{-3}$.  To displace this amount of
material by a distance approximately equal to its length, over the lifetime
of the radio source (assuming $\tau_{\rm rad} \gtrsim 5 \times 10^6$~yr as
a lower limit), would require it to be accelerated to a velocity $\sim
100$~km~s$^{-1}$ over this timescale. Assuming a constant kinetic energy
input, the required kinetic energy flux into the ISM is therefore $F_E \sim
2 \times 10^{41}$~erg~s$^{-1}$. Thus, evacuating the required volume of gas
over the source lifetime would require a conversion efficiency of order
$\lesssim 1$\% of the total jet kinetic power, well within the bounds of
feasibility.

The lack of strong line emission from the jet could suggest a scenario in
which the surrounding ISM is relatively dense. If the interaction with the
jet only results in slow shocks into the gas ($v_{\rm sh} \lesssim 50 -
100$~km~s$^{-1}$), then little or no optical line emission is produced.
However, the lack of Pa$\alpha$ or H2 suggesting no obvious star formation
argues against this.

We conclude that we may indeed be seeing the channel evacuated by the jet
between the AGN and the first bright X-ray/radio knot, despite the lack of
detected emission from the knot itself.

\vskip 0.5cm
\section{\label{sec:conclusions}Conclusions}

Our HST {\it Wide Field and Planetary Camera 2} and
{\it Near Infrared and Multi Object Spectrograph} observations of Centaurus A
(NGC 5128) have provided several new
insights about the active galactic nucleus. 

Detection of unresolved emission in both visible light and near-IR 
suggests that two different emission mechanisms are required:
a non-thermal component which represents an extrapolation of the X-ray power
law, reddened by $A_V\sim 14$ (consistent with mid-IR and X-ray estimates),
and a thermal component probably caused by emission from hot
dust within 2pc of the nucleus.
These results are in agreement with previous suggestions
that Centaurus A harbors a misaligned BL Lac nucleus.
We have shown that only variability on time scales shorter than
$\sim 2$ months can unambiguously prove the existence
of a non-thermal component in the near-IR.

We have detected the 20pc-scale nuclear disk in the $\FeII\lambda 1.64\mu$m
line which shows a morphology similar to that observed in \PA\
with an \FeII/\PA\ ratio typical of low ionization Seyfert galaxies and LINERs.

We do not see evidence for optical emission from, or star formation
associated with, the radio/X--ray jet.  We do detect a blue linear
structure, aligned with jet and extending from the nucleus to knot A, which
we interpret as a region of relatively low reddening. This feature may
represent a channel in which gas and dust have been swept away by the jet,
consistent with simple energetic arguments.

The data allow us to derive a map of dust extinction, E(B-V), in a
20\arcsec$\times$20\arcsec\ circumnuclear region and reveal a several
arcsecond long dust feature near but below the nucleus, oriented in a
direction transverse to the large dust lane.  This structure may be related
to the bar observed with ISO and SCUBA, as reported by Mirabel et al.
(\cite{iso}).
We find rows of Pa$\alpha$ emission knots along the top and
bottom edges of the bar, which we interpret as star formation regions,
possibly caused by shocks driven into the gas. 
The inferred star formation rates are moderately high
($\sim 0.3\,$M$_\odot$ yr$^{-1}$).
If we hypothesize, with
Mirabel, that the bar represents a mechanism for transferring gas in to the
center of the galaxy, then the large dust lane across the galaxy, the bar,
the knots, and the inner Pa$\alpha$ disk previously reported in Paper II,
all represent aspects of the feeding of the AGN.  Gas and dust
are supplied by a recent galaxy merger; a several arcminute-scale bar
allows the dissipation of angular momentum and infall of gas toward the
center of the galaxy; subsequent shocks trigger star formation; and the gas
eventually accretes onto the AGN via the 20pc disk.

Reconstructing the radial light profile of the galaxy to within 0\farcs1 of
the nucleus by combining V, H and K measurements, corrected for reddening,
shows that Centaurus A has a core profile, according to the classification
of Faber et al. (\cite{faber}). Using the models of
van der Marel (\cite{marel}), we
estimate a black--hole mass of $\sim 10^9$M$_\odot$, consistent with ground
based kinematical measurements (Israel \cite{israel}).  This is
consistent with the presence of a strong AGN, as suggested by the large
radio lobes, the jet, and the strong X-ray emission. It further suggests
that the ionized gas disk seen in Pa$\alpha$ (Paper II) would
have relatively high rotational velocities, of order 800 km/sec.  Planned
high spatial resolution near-IR spectroscopy should be able to accurately
determine the mass of the super-massive black hole in this nearest AGN.

\acknowledgments
A.M. acknowledge the partial support of the Italian Space Agency (ASI)
through the grant ARS--98--116/22 and of the Italian Ministry for University
and Research (MURST) under grant Cofin98-02-32.
A.M. and A.C. acknowledge support from the STScI Visitor Program.
A.K. acknowledge support through GO grants O0570/GO-7267
and O0568/GO-6578 from Space Telescope 
Science Institute, which is operated by the Association
of Universities for Research in Astronomy, Inc., under NASA contract
NAS 5--26555.
We thank Nino Panagia, Tino Oliva, Marco Salvati, Roberto Maiolino
and Lucia Pozzetti for helpful discussions which greatly improved this
paper.  We also thank Felix Mirabel and Dave Saunders for useful
discussions and pre-publication access to the ISO data.

\clearpage
\clearpage


\clearpage
\begin{thebibliography}{}
%
\bibitem[1997]{alonso}
Alonso-Herrero A., Rieke M.J., Rieke G.H., Ruiz M., 1997 ApJ, 482, 747
%
\bibitem[1992]{athanas92a}
Athanassoula E., 1992a MNRAS, 259, 328
%
\bibitem[1992]{athanas92b}
Athanassoula E., 1992b MNRAS, 259, 345
%
\bibitem[1998]{axon98}
Axon D.J., Marconi A., Capetti A., Macchetto F.D., Schreier E.J., 
Robinson A., 1998, ApJ, 496, L75
%
\bibitem[1954]{baade}
Baade W., Minkowski R., 1954 ApJ, 119, 215
%
\bibitem[1997]{bahcall}
Bahcall J.N., Kirhakos S., Saxe D.H., Schneider D.P., 1997 ApJ, 479, 642
%
\bibitem[1986]{bailey}
Bailey J., Sparks W.B., Hough J.H., Axon D.J., 1986 Nature, 32, 150
%
\bibitem[1982]{balick}
Balick B., Heckman T.M.,  1982 ARA\&A, 20, 431
%
\bibitem[1992]{barnes}
Barnes J.E., Hernquist L.,  1992 ARA\&A, 30, 705
%
\bibitem[1987]{barvainis}
Barvainis R., 1987 ApJ, 320, 537
%
\bibitem[1987]{binney}
Binney J., Tremaine S., 1987, ``Galactic Dynamics'', Princeton University Press
%
\bibitem[1996]{wfpc2}
Biretta J.A, et al., 1996, {\it WFPC2 Instrument Handbook}, Version 4.0
(Baltimore:STScI)
%
\bibitem[1949]{bolton}
Bolton J.G., Stanley G.J., Slee O.B., 1949 Nature, 164, 101
%
\bibitem[1970]{bowyer}
Bowyer C.S., Lampton M., Mack J., 1971, ApJ, 161, L1
%
\bibitem[1983]{burns}
Burns J.O., Feigelson E.D., Schreier E.J., 1983 ApJ, 273, 128
%
\bibitem[1997]{calnica}
Bushouse H., Skinner C.J., MacKenty J.W., 1997, {\it NICMOS Instrument
Sci. Rept.} 97-28 (Baltimore:STScI)
%
\bibitem[1997]{capetti}
Capetti A., Axon D.J., Macchetto F.D., 1997 ApJ, 487, 560
%
\bibitem[1989]{cardelli}
Cardelli J.A., Clayton G.C., Mathis J.S., 1989 ApJ, 345, 245
%
\bibitem[1997]{carollo97}
Carollo C.M., et al., 1997, ApJ, 481, 710
%
\bibitem[1986]{clarke86}
Clarke D.A., Burns J.O., Feigelson E.D., 1986 ApJ, 300, L41
%
\bibitem[1992]{clarke92}
Clarke D.A., Burns J.O., Norman M.L., 1992 ApJ, 395, 444
%
\bibitem[1998]{derobertis}
De Robertis M.M., Yee H.K.C., Hayhoe K., 1998 ApJ, 496, 93
%
\bibitem[1997]{dopita}
Dopita M.A., et al., 1997, ApJ, 490, 202
%
\bibitem[1984]{draine}
Draine B.T., Lee H.M., 1984 ApJ, 285, 89
%
\bibitem[1979]{dufour}
Dufour R.J., van den Bergh S., Harvel C.A., et al., 1979 AJ, 84, 284
%
\bibitem[1997]{faber}
Faber S.M., Tremaine S., Ajhar E.A., et al., 1997 AJ, 114, 1771
%
\bibitem[1981]{feigelson}
Feigelson E.D., Schreier E.J., Delvaille J.P., et al., 1981 ApJ, 251, 31
%
\bibitem[1997]{granato}
Granato G.L., Danese L., Franceschini A., 1997 ApJ, 486, 147
%
\bibitem[1997]{gonzales97}
Gonzales Delgado R.M., Leitherer C., Heckman T.M.,
Cervino M., 1997, ApJ, 483, 705
%
\bibitem[1998]{gonzales98}
Gonzales Delgado R.M., Heckman T.M., Leitherer C., et al., 1998, ApJ, 505, 174
%
\bibitem[1979]{graham}
Graham J.A., 1979 ApJ, 232, 60
%
\bibitem[1975]{grindlay75}
Grindlay J.E., Schnopper H., Schreier E.J., Gursky H., Parsignault D.R.,
1975, ApJ, 201, L133
%
\bibitem[1986]{heckman86}
Heckman T.M., Smith E.P., Baum S.A., et al., 1986 ApJ, 311, 526
%
\bibitem[1997]{heckman}
Heckman T.M., Gonzales Delgado R.M., Leitherer C., et al., 1997, ApJ, 482, 114
%
\bibitem[1995]{hernquist}
Hernquist L., Mihos J.C., 1995 ApJ, 448, 41
%
\bibitem[1993]{hui}
Hui X., Ford H.C., Ciardullo R., Jacoby G.H., 1993 ApJ, 414, 463
%
\bibitem[1997]{hunt}
Hunt L.K., Malkan M.A., Salvati M., et al., 1997 ApJS, 108, 229
%
\bibitem[1998]{israel}
Israel F.P., 1998 A\&A Rev., 8, 237
%
\bibitem[1996]{jones}
Jones D.L., Tingay S.J., Murphy D.W., et al., 1996 ApJ, 466, L63
%
\bibitem[1971]{kellogg}
Kellogg E., Gursky H., Leong C., Schreier E., Tananbaum H., Giacconi R., 
1971, ApJ, 165, L49
%
\bibitem[1995]{lauer}
Lauer T.R., Ajhar E.A., Byun Y.-I., et al., 1995 AJ, 110, 2622
%
\bibitem[1995]{leitherer}
Leitherer C., Heckman T.M., 1995 ApJS, 96, 9
%
\bibitem[1984]{lepine}
L\'epine J.R.D., Braz M.A., Epchtein N., 1984 A\&A, 131, 72
%
\bibitem[1997]{nicmos}
MacKenty J.W., et al., 1997, {\it NICMOS Instrument Handbook}, Version 2.0
(Baltimore:STScI)
%
\bibitem[1998]{magorrian}
Magorrian J., et al., 1998, AJ, 115, 2285
%
\bibitem[1995]{maiolino95}
Maiolino R., Ruiz M., Rieke G.H., Keller L.D., 1995 ApJ, 446, 561
%
\bibitem[1999]{maiolino99}
Maiolino R., Alonso-Herrero A., Anders S., et al., 1999 ApJ, submitted
%
\bibitem[1983]{malin}
Malin D.F., Quinn P.J., Graham J.A., 1983 ApJ, 272, L5
%
\bibitem[1994]{mihos}
Mihos J.C., Hernquist L.,  1994 ApJ, 425, L13
%
\bibitem[1996]{mihos96}
Mihos J.C., Hernquist L.,  1996 ApJ, 464, 641
%
\bibitem[1999]{iso}
Mirabel I.F., Laurent O., Sanders D.B., et al., 1999 A\&A, 341, 667
%
\bibitem[1988]{moorwood}
Moorwood A.F.M., Oliva E., 1988 A\&A, 203, 278
%
\bibitem[1991]{morganti}
Morganti R., Robinson A., Fosbury R.A.E., di Serego Alighieri S.,
Tadhunter C.N., Malin D.F., 1991 MNRAS, 249, 91
%
\bibitem[1999]{oliva}
Oliva E., Origlia L., Maiolino R., Moorwood A.F.M., 1999 A\&A, submitted
%
\bibitem[1996]{packham}
Packham C., Hough J.H., Young S., et al., 1996 MNRAS, 278, 406
%
\bibitem[1992]{quillen92}
Quillen A.C., de Zeeuw P.T., Phinney E.S., Phillips T.G., 1992 ApJ, 391, 121
%
\bibitem[1993]{quillen93}
Quillen A.C., Graham J.R., Frogel J.A., 1993 ApJ, 412, 550
%
\bibitem[1981]{rieke}
Rieke G.H., Lebofsky M.J.,  1981 ApJ, 250, 87
%
\bibitem[1986]{rodriguez}
Rodriguez-Espinoza J.M., Rudy R.J., Jones B., 1986, ApJ, 309, 76
%
\bibitem[1999]{rxte}
Rothschild R.E., Band D.L., Blanco P.R., et al., 1999 ApJ, 510, 651
%
\bibitem[1989]{sanders}
Sanders D.B., Phinney E.S., Neugebauer G., et al., 1989 ApJ, 347, 29
%
\bibitem[1979]{schreier79}
Schreier E.J., Feigelson E., Delvaille J., et al., 1979 ApJ, 234, L39
%
\bibitem[1981]{schreier81}
Schreier E.J., Burns J.O., Feigelson E.D., 1981 ApJ, 251, 523
%
\bibitem[1996]{schreier96}
Schreier E.J., Capetti A., Macchetto F., Sparks W.B., Ford H.J.,
1996 ApJ, 459, 535 (Paper~I)
%
\bibitem[1998]{schreier98}
Schreier E.J., Marconi A., Axon D.J., Caon N., Macchetto D., Capetti A., 
Hough J.H., Young S., Packham C., 1998, ApJ, 499, L143 (Paper~II)
%
\bibitem[1998]{simpson}
Simpson C., Meadows V.,  1998 ApJ, 505, L99
%
\bibitem[1998]{steinle}
Steinle H., Bennett K., Bloemen H., et al., 1998 A\&A, 330, 97
%
\bibitem[1972]{toomre}
Toomre A., Toomre J., 1972 ApJ, 178, 623
%
\bibitem[1992]{turner}
Turner P.C., Forrest W.J., Pipher J.L., Shure M.A., 1992 ApJ, 393, 648
%
\bibitem[1999]{marel}
van der Marel R.P., 1999 AJ, 117, 744
%
\bibitem[1997]{datahand}
Voit, M., et al., 1997, {\it HST Data Handbook - Vol. I}, Version 3.0
(Baltimore:STScI)
%
\bibitem[1996]{walker}
Walker I.R., Mihos J.C., Hernquist L., 1996 ApJ, 460, 121
%
\bibitem[1993]{weil}
Weil M.L., Hernquist L., 1993 ApJ, 405, 142
%
\bibitem[1995]{photflam}
Whitmore B., 1995, In {\it Calibrating HST: Post Servicing Mission}, 
A. Koratkar and C. Leitherer (Eds.), Baltimore STScI, p.269
%
\bibitem[1999]{wills}
Wills B.J., in ``Quasars and Cosmology'', A.S.P. Conference Series (1999),
eds. G.Ferland, J.Baldwin (San Francisco:ASP), in press (astro-ph/9905093)
%
\end{thebibliography}
\end{document}